\documentclass[pra,showpacs,preprintnumbers,amsmath,amssymb,tightenlines,twocolumn,superscriptaddress]{revtex4}

\usepackage{graphicx}
\usepackage{dcolumn}
\usepackage{bm}
\usepackage{epsfig}
\usepackage{stmaryrd}
\usepackage{color}
\usepackage{amsmath}
\usepackage{psfrag}

\newcommand{\bra}[1]{\langle\,{#1}\, |}
\newcommand{\ket}[1]{|\,{#1}\,\rangle}

\newcommand{\cref}[1]{chapter~\ref{#1}}

\newcommand{\Cref}[1]{Chapter~\ref{#1}}

\begin{document}

\title{Entangled Wavefunctions from Classical Oscillator Amplitudes}


\author{John~S.~Briggs} 
\email{briggs@physik.uni-freiburg.de}
\author{Alexander Eisfeld}
\email{eisfeld@mpipks-dresden.mpg.de}
\affiliation{Max Planck Institute for the Physics of Complex Systems,
N\"othnitzer Strasse 38, D-01187 Dresden, Germany}

\begin{abstract}
In the first days of quantum mechanics Dirac pointed out an analogy between the time-dependent coefficients of an expansion of the Schr\"odinger equation and the classical position and momentum variables solving Hamilton's equations. Here it  is shown that the analogy can be made an equivalence in that, in principle, systems of classical oscillators can be constructed whose position and momenta variables form time-dependent amplitudes which are \emph{identical} to the complex quantum amplitudes of the coupled wavefunction of an $N$-level quantum system with real coupling matrix elements. Hence classical motion can reproduce quantum coherence.
\end{abstract}

\maketitle

\section{Introduction}
In their first formulation of quantum mechanics, both Schr\"odinger and Dirac were strongly influenced by connections to the Hamiltonian formulation of classical mechanics. Indeed, in one of the very first applications of Schr\"odinger's time-dependent equation (TDSE), Dirac \cite{Dirac_PRSA_114}
indicated a close parallel between the coupled first-order set of equations arising from the TDSE and the coupled first-order  Hamilton equations of classical mechanics. Dirac introduced the time-dependent basis set expansion,
\begin{equation}
\label{eq:basis_expansion}
\ket{\Psi(t)} = \sum_n c_n(t) \ket{\pi_n},
\end{equation}
where the $c_n$ are complex co-efficients and $\ket{\pi_n}$ denotes an arbitrary basis. In the TDSE  (where, for the moment, we put $\hbar=1$) this expansion leads to the set of first-order coupled equations,
\begin{equation}
\label{eq:Schreqns}
i\dot{c}_n  = \sum_m H_{nm} c_m,
\end{equation}
where $H_{nm}$ are the matrix elements of the quantum Hamiltonian. Dirac then remarked that by considering the co-efficients $q_n \equiv c_n$ and $p_n \equiv ic^*_n$ to be canonical variables and assuming a 'Hamiltonian function'
\begin{equation}
\label{eq:HamFunc}
\mathcal{H} = \sum_{nm} c^*_n H_{nm} c_m
\end{equation}
the quantum equations are equivalent to the classical  Hamilton equations
\begin{equation}
\label{eq:Hameqns}
\dot{q}_n = \frac{\partial\mathcal{H}}{\partial p_n}~, \qquad\dot{p}_n = -\frac{\partial\mathcal{H}}{\partial q_n}.
\end{equation}
Note, however, that this is still fully quantum mechanical since the matrix
elements and the amplitudes appearing in the 'classical' Hamiltonian are all
complex objects. Hence, there is no obvious classical counterpart. Dirac also
made a transformation to real variables by using amplitude and phase of the
quantum co-efficients, i.e.\
\begin{equation}
c_n = \sqrt{\rho_n} e^{i\theta_n},
\end{equation}
where $\rho_n = c^*_n c_n$. Although with this transformation the variables are real, the 'classical' Hamiltonian still contains complex quantities and hence does not obviously correspond to any real physical system.

Forty years later, Strocchi  \cite{RevModPhys.38.36} approached the question of quantum/classical equivalence slightly differently. He first formulated classical mechanics in terms of complex variables $z_n = (q_n + ip_n)/\sqrt{2}$, with $p_n$ and $q_n$ real, to give the Hamiltonian equations  (\ref{eq:Hameqns}) in the form,
\begin{equation}
i\dot{z}_n = \frac{\partial\mathcal{H}}{\partial z_n^*},
\end{equation}
and its complex conjugate. Here $\mathcal{H}$ \emph{is} a real  classical Hamiltonian. Then it is remarked that the TDSE coupled equations (\ref{eq:Schreqns}) are equivalent to these classical equations i.e. $c_n = z_n$ if the 'classical' Hamiltonian function is taken as the expectation value of the quantum Hamiltonian i.e.\
\begin{equation}
\mathcal{H} = \bra{\Psi(t)}H\ket{\Psi(t)} = \sum_{nm}c^*_n(t) H_{nm} c_m(t).
\end{equation}
as in the treatment of Dirac.
Although ostensibly time-dependent and complex, it is easy to prove that $\mathcal{H}$ becomes constant and real for Hermitian Hamiltonians. Identifying $c_n$ with $z_n$ and reverting to the real variables $(q_n,p_n)$ one has however, still the apparently complex form quoted by Strocchi \cite{RevModPhys.38.36},
\begin{equation}
\mathcal{H} = \frac{1}{2} \sum_{nm} H_{nm}( q_nq_m + p_np_m - i q_np_m + i p_nq_m).
\end{equation}
Of course, for $H$  Hermitian one does achieve a real 'classical' Hamiltonian
\begin{equation}
\mathcal{H} = \frac{1}{2} \sum_{nm}[ H_{nm}( q_nq_m + p_np_m) + 2\Im (H_{nm})q_np_m]
\end{equation}
and in the special case that all coupling matrix elements are real the Hamiltonian
\begin{equation}
\label{eq:classHam}
\mathcal{H} = \frac{1}{2} \sum_{nm} H_{nm}( q_nq_m + p_np_m),
\end{equation}
which is that of coupled real harmonic oscillators. Note that the coupling is of a very special form in which there is both linear position and momentum off-diagonal coupling with exactly the same coupling strengths. 

The mapping of the TDSE onto classical-like equations up until now has been regarded as something of a curiosity, although it has been used  as a starting point to treat some problems of molecular electronic dynamics semi-classically \cite{meyer):3214,PhysRevA.59.64}. Here we show that systems of coupled classical oscillators corresponding to the Hamiltonian Eq.~(\ref{eq:classHam}) can be realised whose position and momentum variables reproduce \emph{exactly} the time-dependent coefficients of an expansion of the quantum wavefunction. We call these the p\&q-coupled oscillators. Each oscillator plays the role of a state in Hilbert space and the coupling between the oscillators the role of quantum coupling matrix elements.
 Unfortunately, for a general quantum system the corresponding exactly equivalent classical system involves rather complicated coupling schemes between the oscillators. However, we show that in a weak-coupling approximation, simpler sets of coupled oscillators in which only the q-coupling in the Hamiltonian Eq.~(\ref{eq:classHam}) occurs provide an exceedingly good approximation to the exact result. 

In previous papers \cite{BrEi11_051911_,EiBr_ARXIV_}, where we studied energy transfer along dipole-dipole interacting molecules, we called this approximation the 'realistic-coupling approximation' (RCA) since for this case, either classical or quantum, to be realistic it is essential that the oscillators and  quantum entities largely retain their character when coupled i.e.\ the coupling is weak compared to their internal forces. 
Although in this work we consider a general Hamiltonian where the diagonal elements in a chosen basis are not necessarily associated with concrete physical entities and thus there is no {\it a priori} reason for the off-diagonal elements to be small, we will still keep the term RCA for this weak coupling approximation.

 It is interesting that, with reference to specific systems, several authors have already recognised the similarity between q-coupled classical  oscillators and  few-level quantum systems. Indeed, already in one of the the first quantum treatments of resonant electronic energy transfer between identical atoms, Frenkel  in 1930 \cite{Fr30_198_} remarked on the essential equivalence of the quantum treatment to the earlier 1925 classical treatment of Holtsmark \cite{Ho25_722_}. In particular he showed that the normal mode frequencies of the classical coupled equations reduce to the quantum eigenvalues in the approximation that we call the RCA. 

 In 1977 McKibben \cite{Mc77_1022_} constructed a system of three coupled mechanical pendula and showed that, in a weak-coupling approximation similar to our RCA, their equations of motion become equivalent to the quantum equations describing the operation of a spin filter on three levels of the hydrogen atom. Later Hemmer and Prentiss \cite{HePr88_1613_} used the same 3-pendulum classical system to interpret  the quite different 3-level atomic problem of  the stimulated resonance Raman effect and pointed out "the strong mathematical similarities between the pendulum amplitude equations and the resonance Raman equations". More recently, Marx and Glaser \cite{MaGl03_338_} have derived an exact correspondence (in the RCA) between the dynamics of three isotropically coupled spins and those of three coupled classical oscillators. Here the Liouville equation for the density matrix and spin correlation functions derived from it were considered.

 In a rather different application, Jolk et~al.~\cite{Jo98_397_} and Kovaleva et~al.~\cite{KoMaKo11_026602_} have pointed out the similarity of the non-crossing behaviour of the eigenfrequencies of a pair of dissimilar  classical oscillators under varying coupling to the non-crossing behaviour of the eigenenergies of two quantum levels subject to a varying perturbation, as in the celebrated Landau-Zener problem. Similarly Spreeuw~et~al.~\cite{SpDrBe90_2642_} has discussed a classical optical system as analogue to a driven two-level atom.

 In previous communications \cite{BrEi11_051911_,EiBr_ARXIV_}  we have shown that the time-dependent Schr\"odinger equation  for an aggregate of $N$ coupled monomers having a single electronic transition and the classical equations for coupled electric dipoles are equivalent in the RCA. There we demonstrated that the quantum coherence in the  transfer of electronic excitation along a linear chain of monomers is reproduced by the transfer of electric dipole strength along the equivalent classical array. The equivalence proof was restricted to this special problem of coupled  classical electric dipoles and its relevance for energy transfer in photosynthesis.

 These many and varied applications point to a general property of equivalence of classical oscillators and quantum systems in finite Hilbert spaces. However, none of the above authors  discuss their findings in the context of Dirac and Strocchi's analysis. Also they all use classical oscillators that are q-coupled, i.e. they work only in the RCA and do not recognise that it is possible to construct classical oscillators that \emph{exactly} mimic the quantum system. In the present paper we show how to construct such p\&q-coupled oscillators  and then clarify under what conditions the RCA allows simpler q-coupled sets of oscillators to be used as analogues of coupled-state quantum systems.

The structure of the paper is as follows.
    In Sec.2.1 we derive the classical p\&q-coupled equations whose variables give directly the coefficients of the time-dependent Schr\"odinger wavefunction. We also examine the derivation of the RCA and the meaning of this approximation in the context of  the Dirac and Strocchi mapping of the Schr\"odinger equation to Hamilton's equations.
As specific example, in Section 3  we consider excitation transfer on a quantum aggregate of $N$ coupled monomers  and its classical equivalent of $N$ coupled pendula. Special attention is given to the $N=2$ dimer case whose wavefunction is considered a fundamental example of  quantum entanglement. In this case the quantum (and therefore also the p\&q-coupled classical) solution can be obtained in simple closed from. Then the RCA is tested by comparing the q-coupled classical solution with the p\&q-coupled solution i.e.\ with the \emph{exact} quantum solution. Oscillatory energy transfer on the dimer is shown to correspond classically to the well-known "beating" phenomenon of coupled pendula from whose motion the entangled wavefunction can be extracted.  

The classical analysis is given, as example, for the specific case of coupled undamped vertical pendula. However it is clear that the analysis also applies to other sets of oscillators, particularly classical electrical LC circuits, as shown in appendix \ref{elec_Osci}. It is also demonstrated in appendix \ref{damping} that the inclusion of a velocity-dependent damping term in the classical equations corresponds exactly to the inclusion of damping via a complex eigenvalue in the quantum equations, as used for example in Ref.~\cite{HePr88_1613_}.

\section{Quantum $N$-level Problem and Coupled Classical Oscillators}
\subsection{Exact mapping of quantum to classical motion}

First we will show that the first-order Schr\"odinger equations (\ref{eq:Schreqns}) map simply onto the classical Hamilton equations for a set of coupled classical oscillators.
Since the classical problem of coupled oscillators is almost always formulated using the Newton equations rather than the Hamilton equations we also present the corresponding second-order Newton equations.
 We start with the Strocchi classical Hamiltonian,  Eq.(\ref{eq:classHam}).
The  Hamilton equations (\ref{eq:Hameqns}) give
\begin{equation}
\dot q_n = \sum_{m} H_{nm}p_m\qquad \dot p_n = -\sum_{m} H_{nm}q_m
\end{equation}
Symbolically, writing $q$ and $p$ as vectors and $H$ as a matrix the above equations are
\begin{equation}
\label{eq:hameq}
\mathbf{\dot q} = \mathbf{H} \mathbf{p}\qquad \mathbf{\dot p} = -\mathbf{H} \mathbf{q}
\end{equation}
and formally
\begin{equation}
\mathbf{\ddot q} = \mathbf{H \dot{p}} = -\mathbf{H}^2\mathbf{q}
\end{equation}
which are a set of coupled oscillator equations and can be solved for $\mathbf{q}(t)$ and  $\mathbf{\dot q}(t)$. The momenta at time $t$ can then be calculated from
\begin{equation}
\label{eq:pdef}
\mathbf{p} = \mathbf{H}^{-1} \mathbf{\dot q}
\end{equation}
The set of  complex amplitudes, the vector $\mathbf{z}$, is constructed as
\begin{equation}
\label{eq:zclass}
 \mathbf{z} = \frac{1}{\sqrt2} (\mathbf{q} + i\mathbf{p}).
\end{equation}
From the Hamilton Eqs.~(\ref{eq:hameq}) we have
\begin{equation}
\mathbf{\ddot z} =  -\mathbf{H}^2\mathbf{z}
\end{equation}
Similarly the Schr\'odinger equation (\ref{eq:Schreqns}) is written
\begin{equation}
i\mathbf{\dot{c}} = \mathbf{H}\mathbf{c}
\end{equation}
or
\begin{equation}
\mathbf{\ddot c} =  -\mathbf{H}^2\mathbf{c}
\end{equation}
which is exactly the classical equation (\ref{eq:zclass}). Hence the p\&q-coupled classical equations and the coupled quantum Schr\"odinger equations are identical and in particular they have the same eigenvalues.

To see how to construct a set of real classical oscillators, now we  write the formal solution in terms of the individual oscillator amplitudes.  First  we split off the diagonal term in the Hamiltonian.
This quantum energy $H_{nn}$ (divided by $\hbar$)  will be denoted by $\omega_n$. The off-diagonal elements (divided by $\hbar$) will be denoted by $V_{mn}$ . Then the Hamilton equations give
\begin{equation}
\begin{split}
\dot q_n =&\ \omega_np_n + \sum_{m\neq n}V_{nm}p_m\\
\dot p_n =& -\omega_nq_n - \sum_{m\neq n}V_{nm}q_m
\end{split}
\end{equation}
Forming $z_n =\frac{1}{\sqrt 2}(q_n + ip_n)$ gives
\begin{equation}
i\dot z_n = \omega_nz_n + \sum_{m \neq n}V_{nm}z_m
\end{equation}
which are identical to the quantum Eqs.~(\ref{eq:Schreqns}) by construction.
Taking the time derivative of the $\dot q_n$ equation and substituting for the $\dot p_n$ from the second equation leads to the coupled
second-order equations
\begin{equation}
\label{eq:couplEqs}
\ddot q_n + \omega_n^2q_n = - \left[\sum_{m\neq n}(\omega_n+\omega_m)V_{nm} - \sum_mW_{nm}\right] q_m
\end{equation}
where we defined 
\begin{equation}
W_{nm} = \sum_{m^\prime\neq n,m^\prime\neq m} V_{nm^\prime}V_{m^\prime m}
\end{equation}
This set of classical equations are the p\&q-coupled equations. Although the coupling elements appear quite complicated, for a given set of quantum matrix elements $V_{nm}$ they can, in principle, be solved to obtain $q_n(t)$ and $\dot q_n(t)$. Then the time-dependent momenta must be calculated by matrix inversion as in Eq.~(\ref{eq:pdef}) and the complex $z_n(t)$, equal to the quantum coefficients $c_n(t)$, calculated.

\subsubsection{Hamiltonian in RCA}
To construct a classical analogue to a given quantum system the oscillators must be coupled as in Eq.~(\ref{eq:couplEqs}). However,  when the Hamiltonian involves only positional $q$ couplings, the Hamilton equations simplify to
\begin{equation}
\begin{split}
\dot q_n = & \ \omega_np_n \\
\dot p_n = &-\omega_np_n - \sum_{m\neq n}H_{nm}q_m
\end{split}
\end{equation}
or equivalently
\begin{equation}
\label{eq:st.class}
\ddot q_n + \omega_n^2q_n = - \sum_{m\neq n}\omega_nV_{nm}  q_m,
\end{equation}
which are the q-coupled classical equations.
 If we define real coupling elements $K_{nm} \equiv - \omega_nV_{nm}$  these are the standard coupled equations, for example of a set of linearly-coupled mechanical or capacitatively -coupled electrical oscillators. In Appendix A it is shown that in the RCA, which corresponds to having $V_{nm} \ll \omega_n, \omega_m,\ \forall n,m$, the solutions  of the simple q-coupled equations are a good approximation to the more complicated p\&q-coupled equations ~(\ref{eq:couplEqs})\\
 
\subsection{Eigenfunctions and Eigenmodes}
\label{sec:eigenfunc+modes}
\subsubsection{Quantum result}
In the basis $\ket{\pi_n}$ defined in eq.~(\ref{eq:basis_expansion}) the Hamiltonian defined in (\ref{eq:Schreqns}) is not diagonal in the general case.
To find the eigenfunctions and eigenenergies we solve
\begin{equation}
H\ket{\psi_k}=E_k\ket{\psi_k}
\end{equation}
The eigenstates $\ket{\psi_k}$ can be expressed in the original basis via
\begin{equation}
\label{eq:k-to-n}
\ket{\psi_k} = \sum_{n=1}^{N}  B_{kn} \ket{\pi_n},
\end{equation}
Denoting the diagonal elements $H_{nn}=\bra{\pi_n}H\ket{\pi_n}\equiv \epsilon_n$ and the off diagonal elements by  $H_{nm}=\bra{\pi_n}H\ket{\pi_m}\equiv \mathcal{V}_{nm}$ the co-efficients $B_{kn}$ can be obtained from the coupled set of equations
\begin{equation}
\label{eq:TIquant}
(E_k - \epsilon_n) B_{kn} = \sum_m \mathcal{V}_{nm} B_{km}
\end{equation}

A general time-dependent wavefunction (coherent wavepacket) can be expanded either in the  basis $\ket{\pi_n}$ or in the eigenbasis $\ket{\psi_k}$, i.e.
\begin{equation}
\label{eq:Psit}
\ket{\Psi(t)} = \sum_n c_n(t) \ket{\pi_n} = \sum_k b_k(t) \ket{\psi_k}
\end{equation}
Since the eigenbasis diagonalises the Hamiltonian, the coefficients $b_k(t)$ are given simply by,
\begin{equation}
b_k(t) = A_k \exp\left(-\frac{i}{\hbar}E_kt\right),
\end{equation}
where the time-independent complex coefficients $A_k$ are decided by the initial conditions.
Then using Eq.~(\ref{eq:k-to-n}) in Eq.~(\ref{eq:Psit}) one has the amplitude of state $n$,
\begin{equation}
\label{eq:quantcnt}
c_n(t) = \sum_k A_k B_{kn} \exp\left(-\frac{i}{\hbar}E_kt\right).
\end{equation}
\subsubsection{Exact mapping}
The Newton equations of motion for the exactly equivalent set of classical oscillators are the coupled equations Eqs.~(\ref{eq:couplEqs}).
 The eigenfrequencies of these classical equations are identical to the $E_k$ of the diagonalised quantum problem. 

Since the eigenvalues and therefore the eigenfunctions of the quantum and classical systems are identical, we see that the classical time dependence can be used to construct the quantum wavefunction. The normal modes arising from the diagonalization of Eqs.~(\ref{eq:couplEqs}) are of the form $q_{k} = \beta_k\cos(\Omega_kt + \alpha_k)$ where $\beta_k$ and $\alpha_k$ are real constants.  From these normal modes we derive the velocities
\begin{equation}
\dot q_{k}(t) = -\Omega_k\beta_k\sin(\Omega_kt + \alpha_k),
\end{equation}
Then the  complex classical amplitudes are $z_{k}(t)  = \frac{1}{\sqrt 2}( q_{k}(t) +ip_{k}(t)) =\frac{1}{\sqrt 2}(q_{k}(t)  +(i/\Omega_k)\dot q_{k}(t))$ to obtain the general solution as a sum of normal modes,
\begin{equation}
\label{eq:classznt}
z_n(t) = \sum_kB_{kn}A_k\exp(-i(\Omega_kt)),
\end{equation}
 where we have to set
 the $t = 0$ initial conditions so that $A_k = (\beta_k/\sqrt 2) \exp(i\alpha_k)$ and put $\Omega_k = E_k/\hbar$.  Comparing this result with Eq.~(\ref{eq:quantcnt}) we see that the classical amplitudes $z_n(t)$ reproduce the quantum co-efficients $c_n(t)$ and the classical motion can be used to reconstruct the quantum entangled wavefunction of Eq.~(\ref{eq:Psit}).
 
 \subsubsection{Coupled equations in RCA}
\label{sec:eigenfunc+modesRCA}
  As we show in Appendix A, in the RCA the Eqs.~(\ref{eq:couplEqs}) reduce to become identical in form to the simpler linearly $q$-coupled classical equations ~(\ref{eq:st.class}), which we write,
 \begin{equation}
 \label{eq:CoupClass}
\ddot q_n + \omega_n^2x_n = -\sum_{m\neq n} K_{nm}q_m.
\end{equation}
Now we show that, again in RCA, these equations have eigenfrequencies which closely approximate those of the quantum problem and therefore also the eigenfrequencies of Eqs.~(\ref{eq:couplEqs}).
Substitution of the special eigenmode solution $q_n = C_{kn} \cos{(\Omega_kt)}$ gives
\begin{equation}
\label{eq:TIclass}
( \Omega_k^2 - \omega_n^2)C_{kn}  = \sum_m K_{nm}C_{km}.
\end{equation}

This equation can be written
\begin{equation}
(\Omega_k - \omega_n)C_{kn} =(\Omega_k + \omega_n)^{-1} \sum_m K_{nm}C_{km}.
\end{equation}
Now we make a second time the  realistic coupling approximation (RCA), by considering   that the spread of eigenfrequencies (bandwidth of the dispersion relation) resulting from diagonalisation of Eq.(\ref{eq:TIclass}) is small compared to the mean natural frequency $\omega \equiv \bar\omega_n$ of the oscillators. Similarly, for non-identical oscillators, the width of the $\omega_n$ distribution must be small compared to $\omega$. Then on the r.h.s.\ of Eq.~(\ref{eq:TIclass}) we can approximate $\Omega_k$ and $\omega_n$ by $\omega$ to give,
\begin{equation}
\label{eq:TIquantclass}
(\Omega_k - \omega_n)C_{kn} = \sum_m \frac{ K_{nm}}{2\omega}C_{km}.
\end{equation}
Further, if we make the identification $E_k \equiv \hbar\Omega_k, \ \epsilon_n \equiv \hbar\omega_n$ and $\mathcal{V}_{nm}/\hbar \equiv K_{nm}/(2\omega)$, Eqs.~(\ref{eq:TIquant}) and (\ref{eq:TIquantclass}) are identical and we can put 
in RCA
\begin{equation}
\label{C_kn=B_kn}
C_{kn} = B_{kn}~ \forall~ k,n.
\end{equation}

Note that the RCA is synonymous with the condition that the couplings  $ K_{nm}$ are small compared to $\omega$ and equivalently in the quantum case that all $\mathcal{V}_{nm}$, are small compared to the mean  energy $\epsilon$ of the various eigenvalues $\epsilon_n$ of $H_0$.  Note also that this in no way implies that perturbation theory must be applicable since, as in the examples given below, the analysis applies to the case where all $\epsilon_n$ are degenerate, where perturbation theory is invalid.

\section{Two Simple Examples}
To illustrate the construction of classical oscillator systems which can reproduce entangled wavefunctions, both exactly and approximately in RCA, we consider the quantum problem of $N$ identical two-level systems. Specifically we consider the dynamics when there is exactly one excitation present and take $\ket{\pi_n}$ as the state in which monomer $n$ is excited and all other monomers are in the ground state. This is the exciton model  studied originally by Frenkel.
 Such model quantum systems are of fundamental importance in several areas, for example in quantum computing and in
the modelling of the photosynthetic unit, in addition to applications mentioned in the introduction. Indeed, the quantum dimer of $N=2$ coupled two-level systems (two qubits), is viewed as having a wavefunction which is the simplest example of quantum entanglement. This is the case we discuss first.

\subsection{The quantum dimer problem and two classical coupled oscillators}
To make the problem concrete we will think of a dimer  composed of two identical atoms or molecules each having only a ground and one excited state. 
Note, that we restrict to the subspace where exactly one excitation is present.
Classically this corresponds to two coupled oscillators.
\subsubsection{Quantum result}
The dimer has $+$ and $-$ eigenstates of the form,
\begin{equation}
\label{eq:dimereigen}
\ket{\psi_{\pm}} = \frac{1}{\sqrt 2}(\ket{\pi_1} \pm \ket{\pi_2})
\end{equation}
with eigenenergies $\epsilon_{\pm} = \epsilon \pm \mathcal{V}$, where $\epsilon$ is the monomer transition energy and $\mathcal{V} \equiv \mathcal{V}_{12} = \mathcal{V}_{21}$ is real. The state $\ket {\pi_1}$ has monomer $1$ excited and monomer $2$ in the ground state and correspondingly for $\ket{\pi_2}$. 

We expand a solution of the TDSE as
\begin{equation}
\ket{\Psi(t)} = a_+(t)\ket{ \psi_+}  + a_-(t) \ket{\psi_-}
\end{equation}
The two eigenstates propagate independently in time according to,
\begin{equation}
\label{eq:cplus}
a_\pm(t) = A_\pm  \exp[-(i/\hbar) \epsilon_\pm t] =  A_\pm \exp[-(i/\hbar)( \epsilon\pm \mathcal{V})t]
\end{equation} 

Excitation transfer is described by the initial condition $\ket{\Psi(0)} = \ket{\pi_1}$ which leads to the time-dependence,
\begin{equation}
\label{eq:PSIT}
\ket{\Psi(t)} = c_1(t) \ket{\pi_1} +  c_2(t) \ket{\pi_2}
\end{equation}
with,
$c_{1/2} = \frac{1}{2} \left[\exp{\left(-i\epsilon_+t/\hbar\right)} \pm \exp{\left(-i\epsilon_-t/\hbar\right)}\right].
$
These expressions can be simplified further to give
\begin{align}
\label{eq:c1quant}
c_1(t) = &  \exp[-(i/\hbar) \epsilon t]  \cos [\mathcal{V}t/\hbar]\\
c_2(t) = & - i \exp[-(i/\hbar) \epsilon t]  \sin [\mathcal{V}t/\hbar]\nonumber
\end{align}
which are the exact quantum solutions and describe a periodic transfer of excitation between the two monomers. Note that in forming the density matrix
of coefficients $c^*_i c_j$ the pure phase factor $\exp[-(i/\hbar) \epsilon t] $ disappears.

\subsubsection{Exact mapping}
The mapped Hamilton equations, with $\omega =\epsilon/ \hbar$ and $V = \mathcal{V}/\hbar$,  for the case $N=2$ are 
\begin{equation}
\begin{split}
\dot q_1 &= \omega p_1 + Vp_2 \qquad \dot p_1 = -\omega q_1 - V q_2\\
\dot q_2 &= \omega p_2 + Vp_1 \qquad \dot p_2 = -\omega q_2 - V q_1
\end{split}
\end{equation}
to give the coupled oscillator equations
\begin{equation}
\begin{split}
\label{eq:Exmap}
\ddot q_{1/2} + (\omega^2 + V^2)q_{1/2}& = -2\omega Vq_{2/1}
\end{split}
\end{equation}
In the usual way these symmetric equations can be diagonalised by the transformation $q_{\pm} = q_1 \pm q_2$ to give normal modes
\begin{equation}
\begin{split}
\label{eq:q_{pm}}
\ddot q_{\pm} + (\omega \pm V)^2q_{\pm}& = 0
\end{split}
\end{equation}
with eigenfrequencies $\Omega_{\pm} = \omega \pm V$,  where we take the positive square root and assume $|V| < \omega$. 
As they should, these reproduce exactly the eigenenergies $\epsilon_{\pm} = \epsilon \pm \mathcal{V}$ of the quantum dimer problem.

The momenta are obtained from the Hamilton equations as
\begin{equation}
\begin{split}
p_{1/2}& = \frac{\omega}{(\omega^2 - V^2)} \left(\dot q_{1/2} - \frac{V}{\omega} \dot q_{2/1}\right)
\end{split}
\end{equation}
which combine conveniently to give
\begin{equation}
p_\pm = p_1 \pm p_2 = \frac{1}{\Omega_\pm} \dot q_\pm.
\end{equation}
Thus in the eigenmodes we have the simple expression for the quantum amplitudes in terms of the classical amplitudes
\begin{equation}
c_\pm = z_ \pm = q_\pm + \frac{i}{\Omega_\pm}\dot q_\pm
\end{equation}
Note that all the above results are invariant to a change of sign of $V$ which only serves to flip the $\Omega_\pm$ eigenvalues.

As example, in the beating mode the wavefunction is exactly reproduced by the classical amplitudes. The initial conditions for beating are $q_1(0) = \beta, \ q_2(0) = 0, \dot q_1 = \dot q_2 = 0$ leading to amplitudes
\begin{equation}
\begin{split}
q_{1/2}(t) &= (\beta/2)[\cos{(\Omega_+t)} \pm \cos{(\Omega_-t)}]
\end{split}
\end{equation}
Forming the velocities and from them the momenta, after some algebra one finds
$p_{1/2}(t) = -(\beta/2)[\sin{(\Omega_+t)} \pm \sin{(\Omega_-t)}]$
so that 
\begin{equation}
\begin{split}
z_{1/2} =& (\beta/2\sqrt 2)[ e^{-i\Omega_+t} \pm e^{-i\Omega_-t}].
\end{split}
\end{equation}
Choosing $\beta = -\sqrt 2$ and noting that $\epsilon_\pm = \hbar\Omega_\pm$,  these are exactly the quantum amplitudes $ c_1, c_2$ of Eq.~(\ref{eq:PSIT}) leading to the explicit beating forms of Eqs.~(\ref{eq:c1quant}).

We now show how to simulate the quantum equation of motion using two coupled classical pendula.
 The classical equations of motion for two (mathematical, i.e\ linearized) pendula with oscillation angle $\phi$, natural frequency $\omega$  and coupled by a spring with coupling strength $K$ are
\begin{equation}
\label{eq:phieqns}
\begin{split}
&\ddot \phi_1 + \omega^2 \phi_1 + K\phi_1 = K \phi_2\\&
\ddot \phi_2 + \omega^2 \phi_2 + K\phi_2 = K \phi_1
\end{split}
\end{equation}
Making again the identification $V = K/(2\omega)$ the equations (\ref{eq:Exmap}) which map exactly the Schr\"odinger equation, can be written in  the same form
\begin{equation}
\begin{split}
\label{eq:St.Osc}
&\ddot q_1 + \omega_s^2q_1 + Kq_1 = Kq_2\\&
\ddot q_2 + \omega_s^2q_2 + Kq_2 = Kq_1
\end{split}
\end{equation}
where
\begin{equation}
\omega_s^2 \equiv \omega^2 - K +K^2/(4\omega^2)
\end{equation}
so that  in this case $\omega_s = \omega - (K/2\omega) = \Omega_-$.
Thus, one takes two pendula of natural frequency $\omega$. One then couples them with strength $K$ and simultaneously adjusts the lengths to give a new natural frequency $\omega_s$. Then the oscillation amplitudes and velocities of this classical system reproduce \emph{exactly} the complex time-dependent amplitudes of the quantum dimer wavefunction, with  transition energy  $\epsilon = \hbar\omega$ and coupling matrix element $V = K/(2\omega)$. Clearly, as alternative one can leave the natural frequency unchanged as $\omega_s$ and then, for given $K$ infer the transition energy $\omega$ of the equivalent quantum system from $\omega_s$. Hence we have shown how to construct a pair of classical oscillators whose motions reproduce the entangled two-qubit quantum wavefunction time-dependence.

\subsubsection{Dimer in RCA}
Previous works pointing out the equivalence of classical oscillator motion and quantum time-development have not used the exact mapping but rather the simpler standard equations
(\ref{eq:phieqns}) or their equivalent and then invoked the RCA. Next we investigate the accuracy of this approach in the simple dimer case.

The eigenfrequencies of the exact mapping Eqs.~(\ref{eq:St.Osc}) are readily calculated to be $\Omega_+^2 = \omega_S^2$ and $\Omega_-^2 = \omega_S^2 +2K$. With $K = 2\omega V$ this translates to the eigenfrequencies $\Omega_\pm = \omega \pm V$ of Eqs.~(\ref{eq:Exmap}) and of course of the quantum problem. The eigenfrequencies of  the standard Eqs.~(\ref{eq:phieqns}) are $\Omega_+ = \omega$ and $\Omega_- = \sqrt{\omega^2 +2K}$ involving the natural oscillator frequency. However in 
 the realistic coupling approximation (RCA) we expand, as in section \ref{sec:eigenfunc+modes},
\begin{equation}
\label{eq:Omegapprox}
\Omega_- = \sqrt{\omega^2 + 2K} \approx \omega + (K/\omega) = \omega + 2V
\end{equation}
to give eigenfrequency difference $\Omega_- - \Omega_+  = 2V$, as in the exact mapping. This implies that, when the RCA is valid, the solutions to the standard Eqs.~(\ref{eq:phieqns}) will be a good approximation to the exact solution, up to an overall phase caused by the shift in absolute value of the eigenfrequencies. This explains the previously-observed close agreement of quantum solutions and standard (q-coupled) oscillator amplitudes.

\begin{figure}[tp]
\psfrag{time}{$ $}
\psfrag{0pi}[t]{$\phantom{\frac{1}{2}}0$ }
\psfrag{abs2}{$|z_1|^2$}
\psfrag{pi4}[t]{$\frac{1}{4}\pi$}
\psfrag{pi2}[t]{$\frac{1}{2}\pi$}
\psfrag{pi34}[t]{$\frac{3}{4}\pi$}
\psfrag{p}[t]{$\phantom{\frac{1}{2}}\pi\phantom{\frac{1}{2}}$}
\psfrag{pi}[t]{$\phantom{2}\pi\phantom{2}$}
\psfrag{2p}[t]{$2 \pi$}
\psfrag{pi25}[t]{$ $}
\includegraphics[width=4.cm]{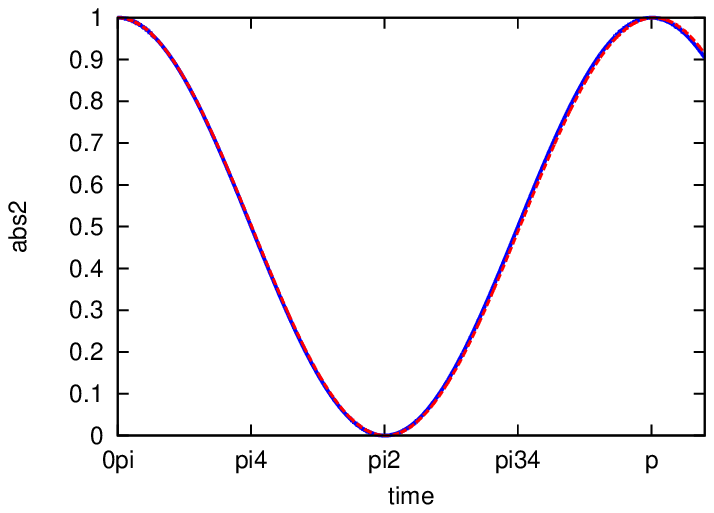}
\psfrag{pi2}[t]{$ $}
\includegraphics[width=4.cm]{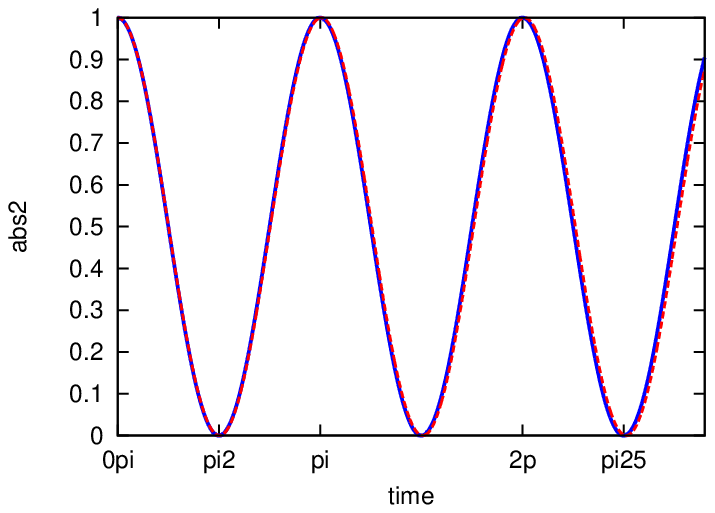}\\
\psfrag{real}{${\rm Re} (z_1)$}
\psfrag{pi2}[t]{$\frac{1}{2}\pi$}
\includegraphics[width=4.cm]{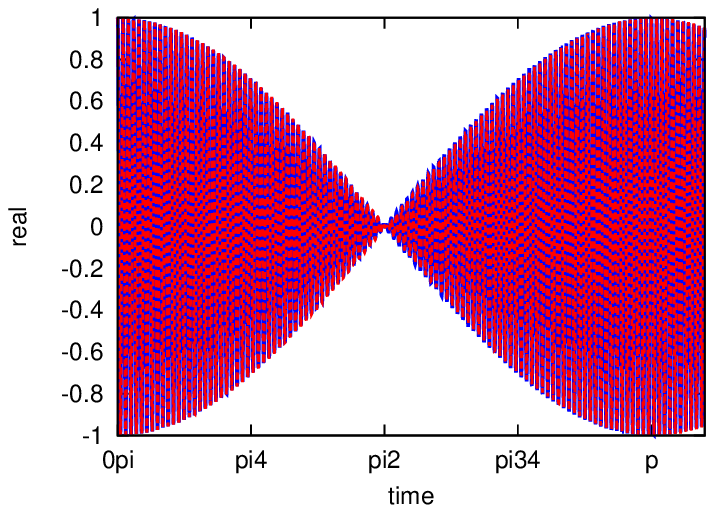}
\psfrag{pi2}[t]{$ $}
\includegraphics[width=4.cm]{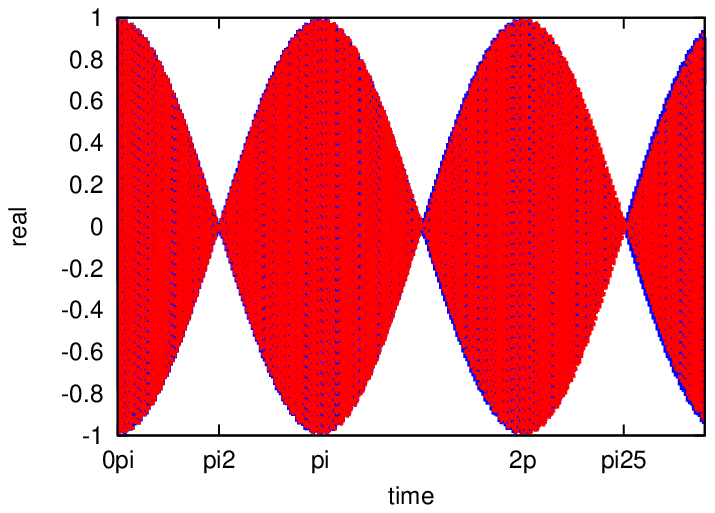}\\
\psfrag{z1z2}{$ z_1^* \, z_2$}
\psfrag{pi2}[t]{$\frac{1}{2}\pi$}
\psfrag{time}[t]{\begin{minipage}{4cm}\vspace{0.2cm} \rm time in $K/(2\omega)$\end{minipage}}
\includegraphics[width=4.cm]{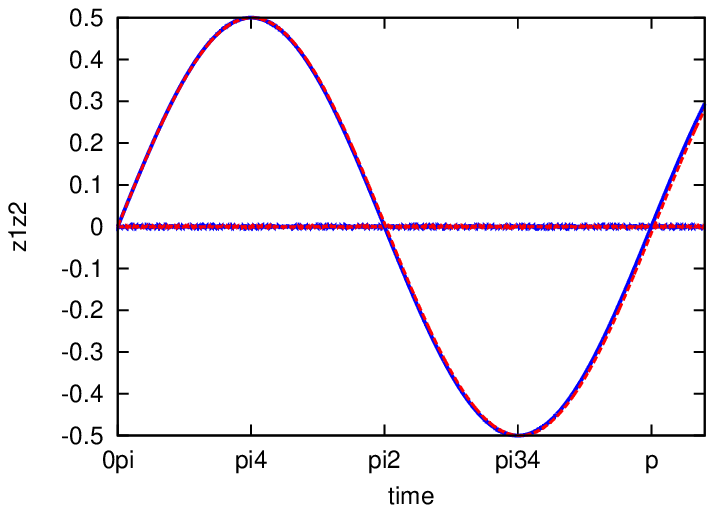}
\psfrag{pi2}[t]{$ $}
\includegraphics[width=4.cm]{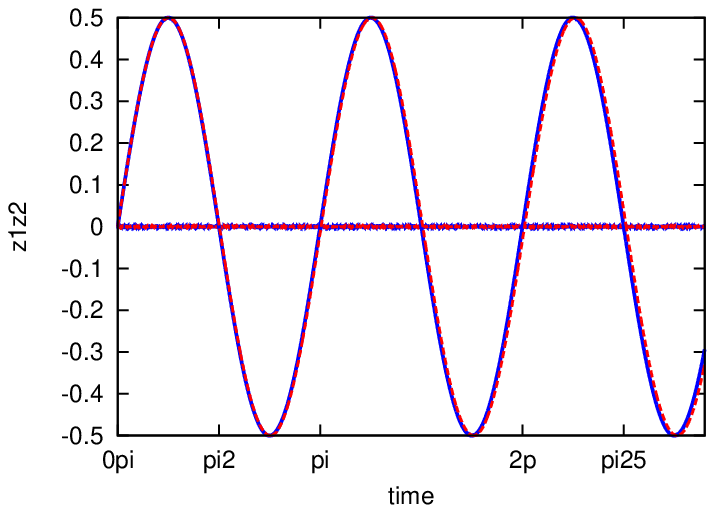}\\
\vspace{0.3cm}
\caption{\label{fig:classical} Time dependence of the quantum or p\&q-coupled classical (blue) and  q-coupled classical (red, dashed) motion. The left column shows details for short times and the right column is for longer times.
The upper row shows the absolute value squared and the middle row the real part of the amplitudes. The bottom panel shows $c^*_1c_2$ and $z^*_1z_2$.
The time is given in units of $K/(2\omega)$ or equivalently $\mathcal{V}/\hbar$}.
\end{figure}

\begin{figure}[tp]
\psfrag{time}{$ $}
\psfrag{0pi}[t]{$\phantom{\frac{1}{2}}0$ }
\psfrag{abs2}{$|z_1|^2$}
\psfrag{pi4}[t]{$\frac{1}{4}\pi$}
\psfrag{pi2}[t]{$\frac{1}{2}\pi$}
\psfrag{pi34}[t]{$\frac{3}{4}\pi$}
\psfrag{p}[t]{$\phantom{\frac{1}{2}}\pi\phantom{\frac{1}{2}}$}
\psfrag{pi}[t]{$\phantom{2}\pi\phantom{2}$}
\psfrag{2p}[t]{$2 \pi$}
\psfrag{pi25}[t]{$ $}
\includegraphics[width=4.cm]{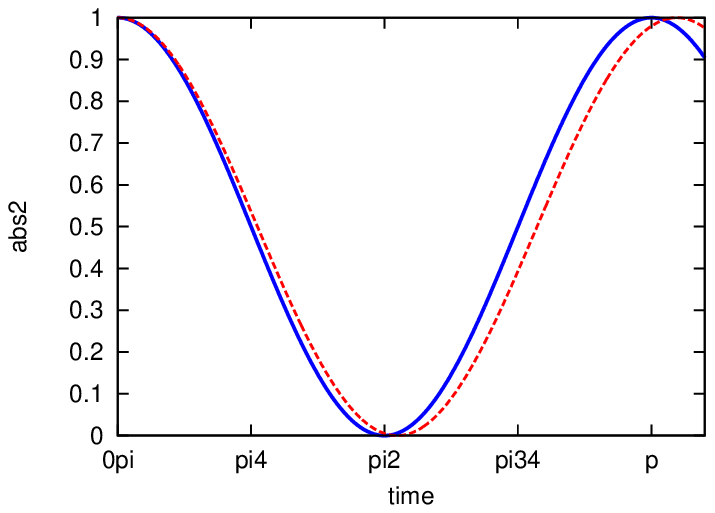}
\psfrag{pi2}[t]{$ $}
\includegraphics[width=4.cm]{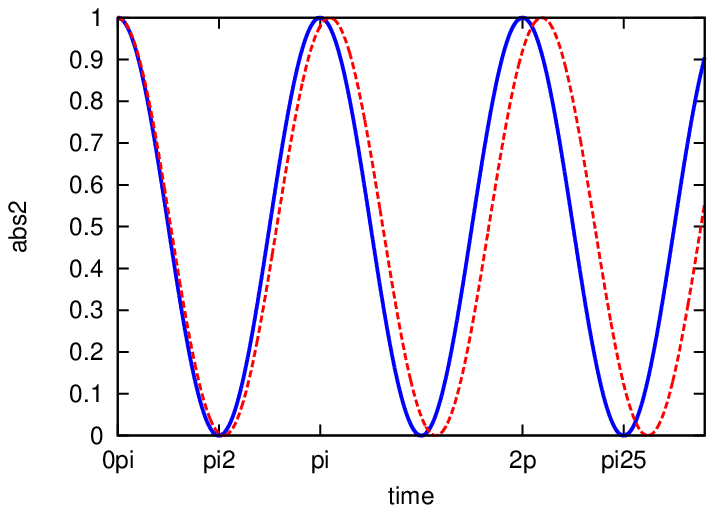}\\
\psfrag{real}{${\rm Re} (z_1)$}
\psfrag{pi2}[t]{$\frac{1}{2}\pi$}
\includegraphics[width=4.cm]{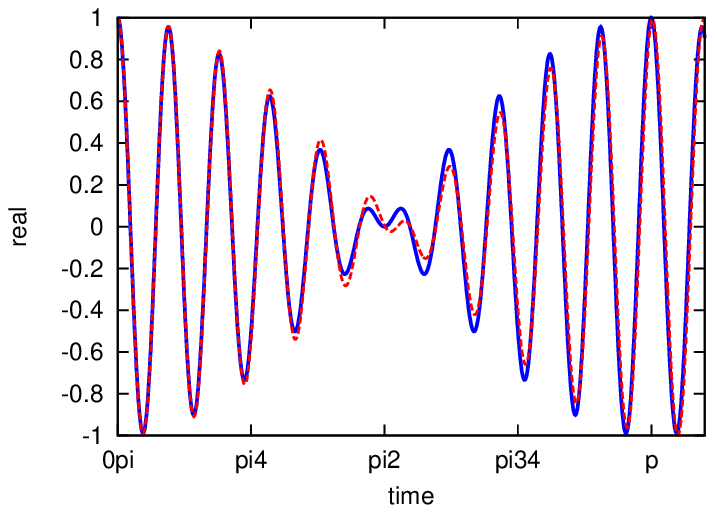}
\psfrag{pi2}[t]{$ $}
\includegraphics[width=4.cm]{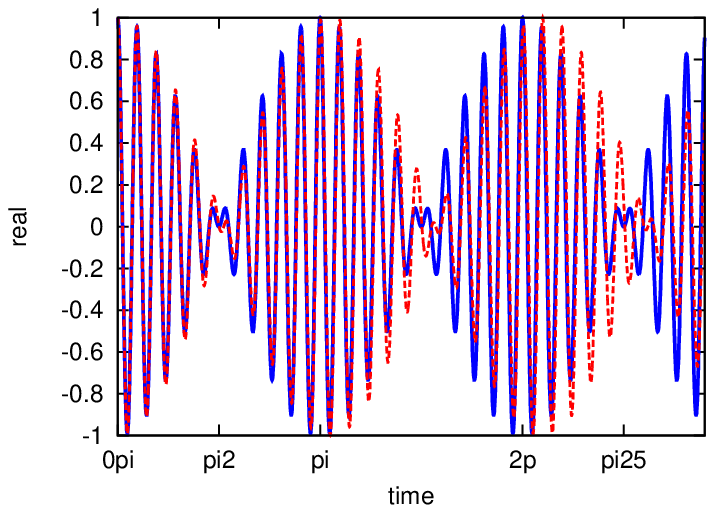}\\
\psfrag{z1z2}{$ z_1^* \, z_2$}
\psfrag{pi2}[t]{$\frac{1}{2}\pi$}
\psfrag{time}[t]{\begin{minipage}{4cm}\vspace{0.2cm} \rm time in $K/(2\omega)$\end{minipage}}
\includegraphics[width=4.cm]{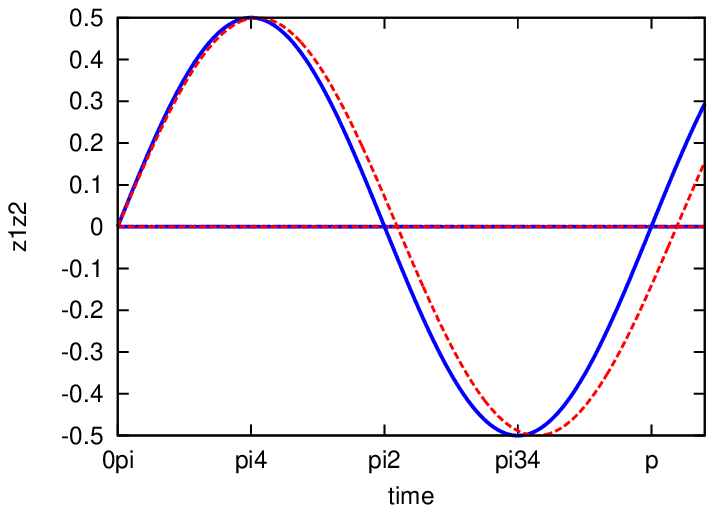}
\psfrag{pi2}[t]{$ $}
\includegraphics[width=4.cm]{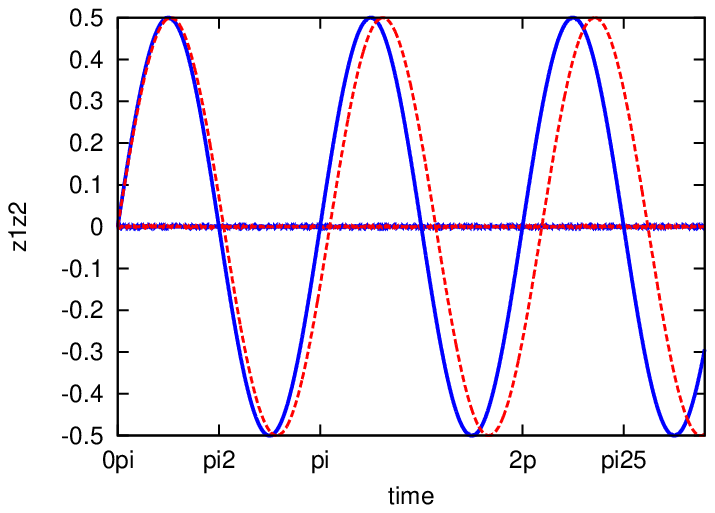}\\
\vspace{0.3cm}
\caption{\label{fig:classical2} Same as Fig.1 except that the coupling strength is $K = 0.1$. } 
\end{figure}

To assess the accuracy of the RCA we have calculated exact quantum (exactly equivalent to the solution of the p\&q-coupled classical Eqs.~(\ref{eq:Exmap})) and compared to solutions of the standard q-coupled classical Eqs.~(\ref{eq:phieqns}) for the $N=2$ dimer case with "beating" initial conditions. The results are shown in Figs.~\ref{fig:classical} and \ref{fig:classical2}. In Fig.~\ref{fig:classical}  the case $\omega = 1,\  K= 0.01$ is shown and one sees excellent agreement between exact, p\&q- coupled (blue, solid) and  standard q-coupled (red, dashed) results over  several periods of the transfer time (the time is given in units of $K/(2\omega)$ or equivalently $\mathcal{V}/\hbar$). Only a small shift in relative phase is perceptible. In the middle column the q-coupled results are multiplied by an overall phase factor $\exp(i(\Omega_- -\omega)t)$ to compensate for the overall energy shift between classical and quantum eigenvalues.  The upper figures show the absolute values squared of the classical q-coupled and quantum coefficients where this phase difference disappears. Similarly the lower panel shows the imaginary part of the off-diagonal density matrix elements $c^*_1c_2$ and $z^*_1z_2$ where the overall phase factor cancels also (the real part is zero).  By contrast to the good agreement for $K=0.01$,  the case $K=0.1$ is shown in Fig.~\ref{fig:classical2}, where one sees the beginning of the breakdown of the RCA in that, although still of the same shape, the phase difference between quantum and classical q-coupled curves is becoming more pronounced.

\subsection{A circular array of $N$ interacting monomers}

For $N=2$ we have shown that by adjusting the natural oscillator frequency to be $\omega_S$ one can readily construct a classical system to mimic exactly the quantum wavefunction. To illustrate the complexity of the equivalent classical system that begins to arise for quantum systems of larger Hilbert space dimension, we consider next the extension to $N$ interacting quantum monomers. This is a model for Frenkel exciton energy transfer in molecular crystals \cite{Me58_647_}, on dye aggregates \cite{MaKue00__,RoScEi09_044909_} or in the photosynthetic unit \cite{AmVaGr00__}. In a previous paper \cite{BrEi11_051911_} we have shown that in the RCA a set of q-coupled classical oscillators gives energy propagation characteristics indistinguishable from the exact quantum result.

\subsubsection{Quantum result}
For simplicity we take the monomers to be identical with transition energy $\epsilon$. Furthermore we take only nearest-neighbour interaction into account and set $\mathcal{V}_{n,n\pm1} \equiv \mathcal{V}$. For circular boundary conditions the result of diagonalisation is standard \cite{MaKue00__}  and gives
\begin{equation}
\label{eq:disperseQuant}
E_k = \epsilon + 2\mathcal{V}\cos k,
\end{equation}
with $k = (2\pi/N)j ,~ j = 0....N-1$. The transformation matrix elements are
\begin{equation}
B_{kn} = \frac{1}{\sqrt N} \exp(ikn).
\end{equation}
 If we consider the time development beginning with only monomer $0$ excited, i.e. $c_n(0) = \delta_{n0}$, then from the orthogonality of normal modes one has
 \begin{equation}
 \label{eq:cntransfer}
c_n(t) = \frac{1}{N}\exp\left(-\frac{i}{\hbar}\epsilon t\right) \sum_k \exp(ikn) \exp\left(-\frac{i}{\hbar}( 2\mathcal{V}\cos k)t\right).
\end{equation}
\subsubsection{Exact mapping}
The exact mapping of this quantum system to the Hamilton equations gives the p\&q-coupled Eqs.~(\ref{eq:couplEqs}) in the form
\begin{equation}
\ddot q_n + (\omega^2 + 2V^2)q_n = -2V\omega(q_{n+1} + q_{n-1}) - V^2(q_{n+2} + q_{n-2})
\end{equation}
Note that, although the first-order Schr\"odinger equations involve nearest-neighbour couplings, the equivalent second-order Newton equations explicitly contain next-nearest neighbour couplings.
Substituting the trial solution $q_{nk} = A_{kn} \exp{(i\Omega t + ikn)}$ gives the eigenfrequency equation 
\begin{equation}
\Omega^2 - \omega^2 -2V^2 = 4V\omega\cos k + 2V^2\cos{2k}
\end{equation}
with solutions
\begin{equation}
\Omega_k = \omega + 2V\cos k
\end{equation}
in agreement with the quantum result  Eq.~(\ref{eq:disperseQuant}). Hence for this  $N$-monomer  case we see again how to construct a system of classical oscillators to reproduce \emph{exactly} the quantum results. However, the couplings between oscillators is becoming more complicated.

\subsubsection{$N$-mer in RCA}
The simpler q-coupled classical equations are those with nearest-neighbour coupling only, and were solved numerically in \cite{BrEi11_051911_}  to compare with the quantum energy transfer result. Here we show analytically that in RCA one obtains the same eigenfrequencies and time-dependence as in quantum or p\&q-coupled classical case. 
We consider a ring of identical coupled pendula. The coupling is between adjacent pendula only.
The standard equations of motion are,
\begin{equation}
\ddot\phi_n + \omega^2\phi_n + 2K\phi_n = K(\phi_{n+1} + \phi_{n-1}),
\end{equation}
 leading to the eigenmode equation
 \begin{equation}
[\Omega_k^2 - (\omega^2 + 2K)] B_{kn} = -K (B_{k,n+1} + B_{k,n-1}).
\end{equation}
The eigenvalues are 
\begin{equation}
\Omega_k^2 - (\omega^2 + 2K) = -2K\cos k,
\end{equation}
which, in the RCA becomes
\begin{equation}
\Omega_k \approx (\omega + K/\omega) - (K/\omega)\cos k.
\end{equation}
Now we make the identification  $E_k \equiv \hbar\Omega_k, \epsilon \equiv \hbar\omega$ and $\mathcal{V}/\hbar \equiv -K/(2\omega)$ to give,
\begin{equation}
E_k = (\epsilon - 2\mathcal{V}) + 2\mathcal{V}\cos k.
\end{equation}
Comparison to the quantum dispersion relation Eq.~(\ref{eq:disperseQuant}) shows an overall shift of eigenenergies by $2\mathcal{V}$. This is immaterial as it leads only to an overall phase factor. Hence, as was shown numerically in \cite{BrEi11_051911_} for a particular  set of initial conditions, the q-coupled classical oscillators describe to vey good approximation the exciton dynamics on an $N$-monomer chain.

\section{Conclusions}
 We have shown how the eigenenergies and eigenfunctions of an $N$-level quantum system with real coupling matrix elements can be reproduced by a suitable array of coupled classical oscillators. Hence, by observing the classical motion, one can reconstruct the quantum mechanical time-dependent wavefunction or density matrix. The difficulty of realisation of a classical system mimicking the quantum entanglement, is mainly in the condition that a realistic system must reproduce \mbox{\it all} elements of the quantum couplings faithfully. This will be a task of increasing difficulty as the complexity  of the quantum system increases. Nevertheless, we have shown explicitly that such a parallel is feasible for a circular or linear array of identical monomers. This is realised by a corresponding array of coupled pendula. Similarly, the excitation transfer on a molecular dimer, or similar coupled two-level quantum systems, can be faithfully simulated by the beating motion of a pair of classical oscillators, either mechanical or electrical. Such experiments are carried out routinely in undergraduate physics laboratories.

Realistically the multi-oscillator couplings necessary to construct p\&q-coupled oscillators could be attained more easily with miniature LC oscillators than with mechanical ones.  However,we have shown also that the complexity of p\&q-coupled oscillators can be circumvented by using only q-coupled oscillators in the  RCA. This explains the success of previous comparisons of quantum time-dependent motion with that of physical q-coupled oscillator systems.

 Throughout we have neglected the effects of coupling of the quantum system or
 the classical oscillators to the environment. Of course in real systems such
 de-cohering or dissipating couplings are omnipresent. In the quantum case in
 the form of radiative or non-radiative decay of excited states or in
 eigenenergy shifts. In the classical case in the form of frictional forces
 for pendula or the equivalent resistance in the electrical case. At the
 phenomenological level of simply assigning a complex energy to quantum levels
 or, equivalently, adding a  term proportional to velocity to the classical
 equation of motion, it is easy to show that the quantum/classical equivalence
 is preserved (see Appendix \ref{damping}).  We have undertaken a more detailed study of this equivalence in the framework of
 standard theories of open quantum systems \cite{EiBr_ARXIV_}.  Similarly, the
 correspondence between inducing radiative transitions in quantum systems and driving
 classical oscillators with external fields is under investigation.

\begin{acknowledgements}
 We are grateful to Prof.\ Hanspeter Helm for many helpful discussions and assistance. The observation that the RCA is equivalent to the rotating-wave approximation is due to Prof.\ Gerhard Stock and we are grateful for this insight.
\end{acknowledgements}

\appendix
\section{Commentary upon the RCA}

In the foregoing we have derived the RCA by approximating the classical eigenmode frequencies. Further light can be shed on the nature of this approximation by looking at alternative derivations. A useful variation is to include explicitly the eigenvalue phases in the quantum time development of Eq.~(\ref{eq:Psit}) i.e. to write
\begin{equation}
\Psi(t) = \sum_n a_n(t) e^{-\frac{i}{\hbar}\epsilon_nt} \ket{\pi_n}.
\end{equation}
Forming the equivalent classical Hamiltonian as before we obtain
\begin{equation}
\mathcal{H}(t) = \sum_{nm}H_{nm}a^*_n(t)e^{i\omega_nt}  a_m(t)e^{-i\omega_mt},
\end{equation}
where we have set $\omega_n = \epsilon_n/\hbar$.
Note that if one proceeds to second quantization and elevates the coefficients $a_n$ to being annihilation operators, then this equation in just that of the quantum Hamiltonian in the Heisenberg representation and the time-dependent phase factors are just the time propagators of the creation and annihilation operators. In particular, for fixed $n$ and $m$ the off-diagonal coupling terms are of the form, for $H$ hermitian and real
\begin{equation}
\label{eq:quantform}
\mathcal{H}_{nm} = H_{nm} [a^*_na_m e^{i(\omega_n-\omega_m)t} + a^*_ma_n e^{-i(\omega_n-\omega_m)t}].
\end{equation}
 If one now sets
  \begin{equation}
a_ne^{i\omega_nt} \equiv \frac{1}{\sqrt 2}(q_n(t) + ip_n(t))
\end{equation}
corresponding to classical amplitudes $q_n(t),p_n(t)$ or equivalently Heisenberg operators, one obtains coupling elements
\begin{equation}
\mathcal{H}_{nm} = \frac{1}{2}H_{nm}(q_nq_m + p_np_m)
\end{equation}
which are those of Eq.~(\ref{eq:classHam}). By contrast, if one restricts the coupling to terms $q_nq_m$ as in the q-coupled case, one has
\begin{equation}
\begin{split}
\mathcal{H}_{nm} = H_{nm}[a^*_na_m e^{i(\omega_n-\omega_m)t} + a^*_ma_n e^{-i(\omega_n-\omega_m)t}\\
+ a^*_na^*_m e^{i(\omega_n+\omega_m)t} + a_na_m e^{-i(\omega_n+\omega_m)t}].
\end{split}
\end{equation}
Then one sees that to obtain the hamiltonian of Eq.~(\ref{eq:quantform}) which maps exactly to the Schrodinger equation, it is necessary to neglect the "non-rotating wave"
terms involving the rapidly oscillating phase factors $e^{\pm i(\omega_n+\omega_m)t}$, which contribute weakly to transition probabilities. More exactly, let us take $K$ to be the value of the largest of the elements $H_{nm}$ i.e. $H_{nm} = K\gamma_{nm}$ where all $\gamma_{nm}$ are less than unity. Then we can define a dimensionless time as $\tau = Kt$. The rotating wave factors are then of the order of $\exp(i(\Delta\omega/K)\tau)$, where $\Delta\omega$ is the mean frequency difference and the non-rotating wave terms are of the order of $\exp(i(\omega/K)\tau)$, where $\omega$ is the mean frequency. Now if $\omega/K \gg 1$ and $\Delta\omega/K \approx 1$, which is the RCA, again we see that classical and quantum couplings will be the same in that the non-rotating wave terms can be neglected.\\
A rather different view of the RCA is obtained from the Hamilton equations. The resulting q-coupled Newton equations are
\begin{equation}
\ddot q_n + \omega_n^2q_n = -\omega_n \sum_{m\neq n}H_{nm}q_m
\end{equation}
 which are identical in form to Eq.~(\ref{eq:CoupClass}) for standard classical oscillators. By contrast, for the p\&q-coupling one has 
\begin{equation}
\ddot q_n + \omega_n^2q_n = -\omega_n\sum_{m\neq n}H_{nm}q_m - \sum_{m\neq n}H_{nm}\sum_{m^\prime\neq m}H_{mm^\prime} q_{m^\prime},
\end{equation}
which differ from the Newton equations by the last term. However, again scaling as above we have
\begin{equation}
\ddot q_n + \left(\frac{\omega_n}{K}\right)^2q_n = -\frac{\omega_n}{K}\sum_{m\neq n}\gamma_{nm}q_m - \sum_{m\neq n}\gamma_{nm}\sum_{m^\prime\neq m}\gamma_{mm^\prime} q_{m^\prime}.
\end{equation}
Since in RCA we have $\frac{\omega_n}{K}\gg 1$ we can neglect the second order term so that the p\&q-coupled hamiltonian gives the same second-order classical equations of motion as the q-coupled hamiltonian.

\section{Coupled Electrical Oscillators}
\label{elec_Osci}
As above, for the moment we ignore friction of resistance and consider an $LC$ oscillator circuit. For a single oscillator the balance of e.m.f. from inductance and capacitor is expressed by,
\begin{equation}
-L\frac{dI}{dt} + \frac{q}{C} = 0,
\end{equation}
where $q$ is the charge and $I$ is the current related by 
\begin{equation}
I = -\frac{dq}{dt}.
\end{equation}
Then we have the harmonic oscillator equation
\begin{equation}
L\ddot q+ \frac{q}{C} = 0
\end{equation}
or,
\begin{equation}
\ddot q + \omega^2 q = 0
\end{equation}
where the frequency is $\omega = \sqrt{1/LC}$. Note that, compared to a mechanical oscillator, $L$ plays the role of mass, $1/C$ that of coupling constant and $-LI$ corresponds to momentum.

Now we couple two identical $LC$ oscillators by a capacitor $C_K$  connected by leads in which current $J = \frac{dQ}{dt}$ flows. The equations of the coupled circuits are,
\begin{equation}
\begin{split}
&-L\frac{d(I_1 - J)}{dt} + \frac{q_1}{C} = 0,\\&
-L\frac{d(I_2 + J)}{dt} + \frac{q_2}{C} = 0,
\end{split}
\end{equation}
with the e.m.f. balance
\begin{equation}
\frac{q_1}{C} = \frac{q_2}{C} - \frac{Q}{C_K}.
\end{equation}
Using $J = - \frac{dQ}{dt}$ one can eliminate $J$ to obtain,
\begin{equation}
\begin{split}
&(1 + K)\ddot q_1 + \omega^2q_1 - K\ddot q_2 = 0\\&
(1 + K)\ddot q_2 + \omega^2q_2 - K\ddot q_1 = 0,
\end{split}
\end{equation}
where the dimensionless ratio $K \equiv C_K/C$ is defined. These equations can be put also in the form of Eqs.~(\ref{eq:phieqns}),
\begin{equation}
\begin{split}
&(1 + 2K)\ddot q_1 + (1 + K)\omega^2q_1 + K\omega^2q_2 = 0\\&
(1 + 2K)\ddot q_2 + (1 + K)\omega^2q_2 + K\omega^2q_1 = 0.
\end{split}
\end{equation}
Clearly when $K \rightarrow 0$ the equations become those of uncoupled oscillators.
Adding and subtracting either of these two sets of equations leads to uncoupled equations in the new variables $q_{\pm} = q_1 \pm q_2$, i.e.
\begin{align}
\ddot q_+ + \Omega_+^2q_+ =& 0\\
\ddot q_- + \Omega_-^2q_- =& 0
\end{align}
which are identical to Eqs.~(\ref{eq:q_{pm}}) except that now $\Omega_+ = \omega = \sqrt{1/LC}$ and $\Omega_- = \omega/\sqrt{(1+2K)}$. For coupled $LC$ oscillators the RCA corresponds to $K \ll 1$ to give,
\begin{equation}
\Omega_- \sim \omega - K\omega.
\end{equation}
Comparison with Eq.~(\ref{eq:Omegapprox}) shows that in this case we make the identification $2V/\hbar = K\omega$. Then the coupled electrical oscillator equations giving $q_1(t), q_2(t)$ are identical to the q-coupled pendula equations  and, as we have shown, in RCA reproduce the complex amplitudes of the quantum dimer.

\section{Inclusion of Damping}
\label{damping}
To  include damping phenomenologically in the classical case we add a velocity-dependent term to the oscillator equations (\ref{eq:CoupClass}), ignoring the corresponding fluctuations, i.e.
 \begin{equation}
\ddot x_n +2\Gamma_n \dot x_n + \omega_n^2x_n = -\sum_m K_{nm}x_m.
\end{equation}
Then the coupled equations (\ref{eq:TIclass}) become,
\begin{equation}
( \Omega_k^2 - \omega_n^2 + 2i\Omega_k \Gamma_n)C_{kn}  = \sum_m K_{nm}C_{km}.
\end{equation}
This equation is written in the form,
\begin{equation}
(\Omega_k - \omega_n + 2i\frac{\Omega_k}{(\Omega_k+\omega_n)} \Gamma_n)C_{kn} =\frac{1}{(\Omega_k + \omega_n)} \sum_m K_{nm}C_{km}.
\end{equation}
Again, in the RCA, we consider that the spread in eigenfrequencies $\Omega_k$ and $\omega_n$ is small compared to the mean natural frequency $\omega$ to approximate the above equations by,
\begin{equation}
(\Omega_k - \omega_n + i\Gamma_n) C_{kn} = \frac{1}{2\omega} \sum_m K_{mn} C_{km}
\end{equation}
If we assign each quantum level a width $\gamma_n$ then the quantum coupled equations become,
\begin{equation}
(E_k - \epsilon_n + i\gamma_n)B_{kn} = \sum_m \mathcal{V}_{nm} B_{km}
\end{equation}
which are identical in form to the classical equations.
In Ref.~\cite{EiBr_ARXIV_} it is shown, how pure dephasing can be realized in a system of coupled oscillators which reproduces the quantum results.

\end{document}